# Ultra-wideband integrated microwave photonic multi-parameter measurement system on thin-film lithium niobate

*Yong Zheng, Zhen Han, LiHeng Wang, Pu Zhang, YongHeng Jiang, HuiFu Xiao, XuDong Zhou, Mingrui Yuan, Mei Xian Low, Aditya Dubey, Thach Giang Nguyen, Andreas Boes, Qinfen Hao, Guanghui Ren*, Arnan Mitchell and Yonghui Tian**

Y. Zheng, Z. Han, L.Wang, P. Zhang, Y. Jiang, H. Xiao, X. Zhou, M. Yuan, Y. Tian
School of Physical Science and Technology, Lanzhou University, Lanzhou 730000, Gansu, China
E-mail: tianyh@lzu.edu.cn

M. X. Low, A. Dubey, T. G. Nguyen, G. Ren, A. Mitchell
Integrated Photonics and Applications Centre (InPAC), School of Engineering, RMIT University, Melbourne, VIC 3001, Australia
E-mail: guanghui.ren@rmit.edu.au

A. Boes
School of Electrical and Mechancial Engineering, The University of Adelaide, Adelaide, SA 5005, Australia
Institute for Photonics Advanced Sensing, The University of Adelaide, SA 5005, Australia

Q. Hao
Institute of Computing Technology Chinese Academy of Sciences Beijing 100190, China

***Corresponding authors: guanghui.ren@rmit.edu.au, tianyh@lzu.edu.cn**

## Abstract

Research on microwave signal measurement techniques is risen, driven by the expanding urgent demands of wireless communication, global positioning systems, remote sensing and 6G networks. In stark contrast with traditional electronic-based realization, the implementations of microwave signal measurement systems based on integrated compact photonic chip have exhibited distinct advantages in high operation bandwidth, light weight, and strong immunity to electromagnetic interference. However, although numerous integrated microwave photonic signal measurement systems have been reported, measurement bandwidth of the majority of them is still below 30 GHz due to the bandwidth limitation of electro-optical modulators (EOMs). Furthermore, previous studies often are more focused on the measurement of one single parameter (typically the frequency) of microwave signals, which has hindered their practical application in complex situations. Here, an integrated photonic microwave multi-parameter measurement system composed of microwave frequency measurement module and microwave phase amplitude measurement module based on thin-film lithium niobate (TFLN) platform is reported. Utilizing this system, not only the ultra-high bandwidth (up to 60GHz) of microwave frequency, phase and amplitude measurement with low root-mean-squares errors (450MHz, 3.43° and 1.64% of the measurement for frequency, phase and amplitude, respectively), but also the time-domain reconstruction of sinusoidal microwave signals is achieved. This demonstration further broadens the application of integrated TFLN photonic devices in microwave signal measurement technology to address the bandwidth bottleneck of the ever-growing microwave networks in the future information society.

## Introduction

Since the dawn of the information age in the 21st century, the capability to measure the parameters of microwave signals rapidly and accurately has played an increasingly crucial role in applications ranging from traditional military domains such as electronic warfare and radar detection to nascent civilian domains such as satellite communications and the Internet of Things.[1-3] In the imminent sixth-generation (6G) era, it is not difficult to predict the emergence of a bottleneck in operation bandwidth of microwave signal measurement technology since the expansion of radio-frequency (RF) network fails to match with the increasing size, weight, and power (SWaP) of electronics that support higher operating bandwidth.[4-6] Thanks to the excellent characteristics of high flexibility, strong immunity to electromagnetic interference and high operation bandwidth,[7-8] the microwave signal measurement systems based on integrated photonic technology[9-25] have the potential to solve the aforementioned problems and have been widely reported. However, despite the tremendous progress, the proposed integrated microwave photonic measurement system still exhibits certain deficiencies in aspects. On one hand, the bandwidth of the majority of the reported microwave signal measurement systems based on integrated photonic technology is limited to below 30 GHz due to the finite bandwidth of electro-optical modulators (EOMs). On the other hand, previous works primarily focused on measuring a single parameter of the microwave signals, which has restricted their applicability in complex scenarios.

Recent decades have witnessed the accelerated development of the thin-film lithium niobate (TFLN) platform, which makes it a promising candidate for integrated microwave photonic measurement systems. In one aspect, TFLN platform exhibits the excellent characteristics of low loss, high stability, a wide transparent window, and especially excellent linear electro-optical effect, which align with the requirements of high-performance integrated photonic microwave signal measurement systems.[26-27] In the other aspect, the manufacturing process of the TFLN platform is becoming increasingly refined, which has the effect of reducing the production cost of the TFLN platform and therefore facilitating its large-scale commercialization.[28-30] In recent years, a considerable number of photonic devices[31-41] and systems[17,28,42-44] on the TFLN have been reported, laying the foundation for the practical deployment of integrated photonic signal microwave measurement techniques based on the TFLN in the future.

Here, we demonstrate an ultra-wideband integrated photonic microwave multi-parameter measurement system capable of measuring the frequency, phase and amplitude of microwave signals. Benefiting from the excellent modulation performance of the TFLN platform, the measurement with an ultra-high bandwidth (up to 60GHz) and with a low root-mean-squares error (450MHz, 3.43° and 1.64% of the measurement for frequency, phase and amplitude, respectively) are achieved. In addition, the system is capable of identifying the types of microwave signals, including chirp frequency (CF), hopping frequency (HF), phase hopping (HP), amplitude hopping (AP), and phase-amplitude hopping frequency signals. Furthermore, the reconstruction for the time-domain information of an unknown sinusoidal microwave signal is experimentally realized, as all parameters (frequency, phase and amplitude) of its time-domain characteristic can be measured. This system broadens the way for the application of optical technology based on the TFLN platform to microwave measurement in higher frequency bands, and is well-suited for a multitude range of practical applications including satellite communication and unmanned vehicles that require microwave signal measurement in the upcoming 6G era.

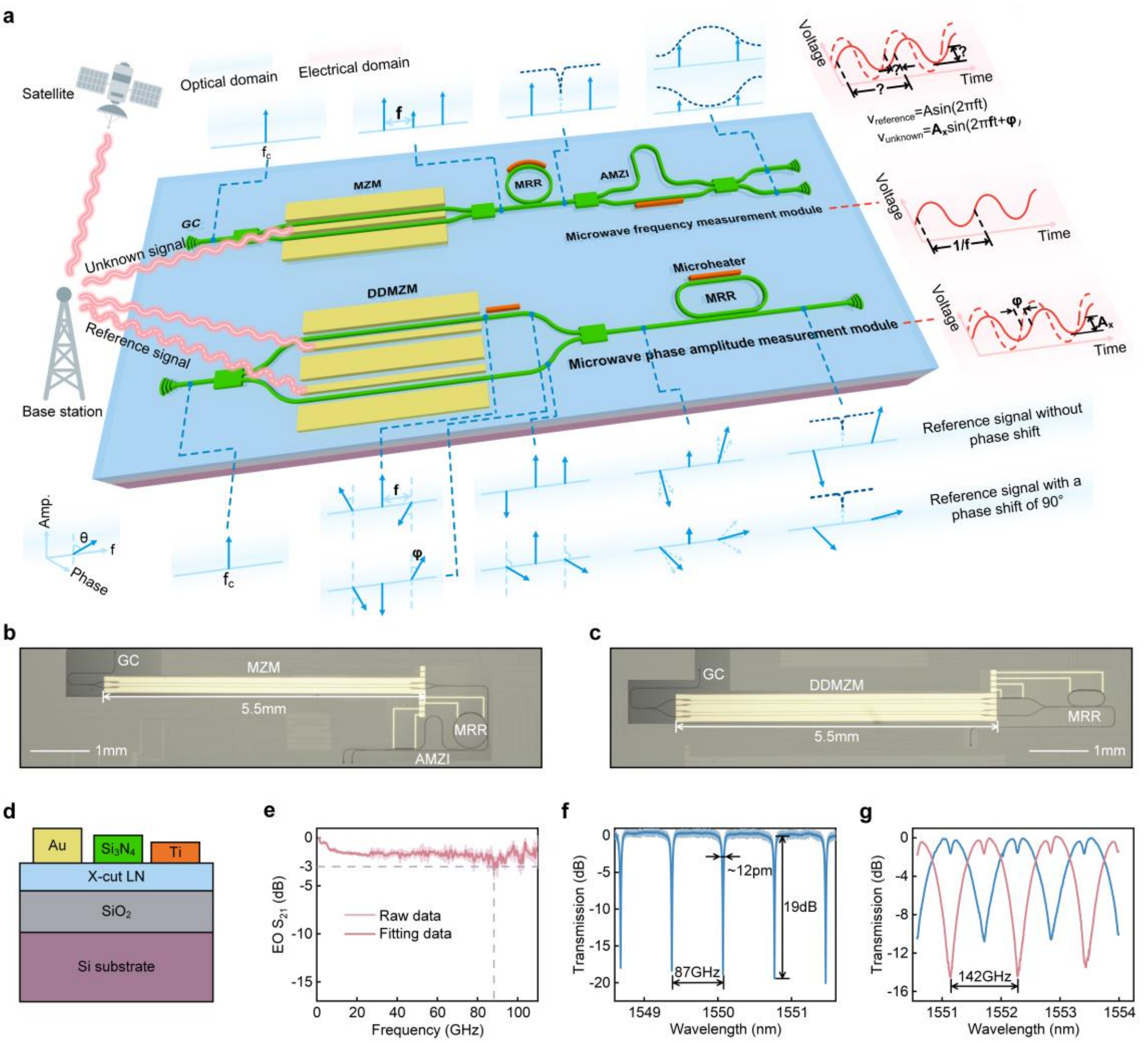

**Figure 1. Integrated microwave photonic multi-parameter measurement system on thin-film lithium niobate a.** Conceptual drawing of the integrated TFLN-based microwave photonic multi-parameter measurement system composed of the microwave frequency measurement module and microwave phase amplitude measurement module. Insets schematically show the optical frequency-domain signal spectra (blue) at different locations of the chip and electrical voltage-time diagrams (pink). The illustration on the left shows an envisioned application scenario where the integrated photonic signal

measurement chip is used for satellite communication. **b-c.** The optical microscope image of the microwave frequency measurement module (b) and the microwave phase amplitude measurement module (c). **d.** The cross-section of the chip. **e.** The EO $S_{21}$ of the MZM. **f.** The transmission spectrum of the MRR. **g.** The transmission spectrum of the AMZI. TFLN: thin-film lithium niobate, MZM: Mach-Zehnder modulator, MRR: micro-ring resonator, AMZI: asymmetrical Mach-Zehnder interferometer, GC: grating coupler, DDMZM: dual-drive Mach-Zehnder modulator.

## Structure and principle

Figure 1(a) presents the conceptual illustration and working principle of our integrated microwave photonic signal measurement system. The system is composed of a microwave frequency measurement module and a microwave phase amplitude measurement system, of which the optical microscope image is shown in Figure 1b-c. The following section will provide a detailed explanation of the composition and operating principle of the two modules.

Microwave frequency measurement module consists of grating coupler (GC), Mach-Zehnder modulator (MZM), micro-ring resonator (MRR), and the asymmetrical Mach-Zehnder interferometer (AMZI). Firstly, a continuous wave optical carrier is generated by a tunable laser source (TLS) and coupled into the chip via an on-chip grating coupler. Then, the unknown signal is modulated onto the optical carrier. Concurrently, the bias voltage must be adjusted to ensure that the MZM is operating at the null transmission point, thereby leading to the generation of a dual sideband suppressed-carrier (DSB-SC) signal with the even order sideband suppressed (see Note S1, Supporting Information for more detail). The MRR followed further suppress the optical carrier without affecting the first-order sideband, which is achieved by tuning its resonant to match with the carrier wavelength. Subsequently, the DSB-SC signal is subjected to filtering by AMZI, which serves as a pivotal linear optical discriminator within the microwave frequency measurement module. Since the two interference arms are not equal in length, the transmission spectra of two output ports of the AMZI exhibit complementary periodic oscillations with the free spectrum range (FSR) being determined by the length difference between the two arms (see Note S2, Supporting Information for more detail). When the center of the transmission spectrum of the AMZI output port is aligned with the optical carrier, a complementary response is observed in the optical power of the two output ports relative to the microwave frequency. The optical power ratio of the two output ports, commonly regarded as the amplitude comparison function (ACF), is a one-to-one correspondence function solely related to the microwave frequency. It establishes the frequency-to-power mapping (FTPM), which allows the frequency of the unknown microwave signal to be measured by directly detecting the direct current (DC) component of the photocurrent.

Microwave phase and amplitude measurement module is composed of GC, dual-drive Mach-Zehnder modulator (DDMZM), and MRR. The optical carrier wave is divided equally into two beams of equal power into the two interfering arms of the DDMZM via the 1×2 multimode interference (MMI). The reference signal with known amplitude is necessary to characterize the phase information of the unknown signal. The reference signal and the unknown signal are modulated onto the optical carrier on the upper and lower interference arms of the DDMZM respectively. The optical signal on one arm of the DDMZM acquires an additional 180° phase shift due to the heating of the micro heater, and subsequently interferes with the optical signal on the other arm after passing through 2×1 MMI .The power of optical carrier is significantly diminished as a consequence of the destructive interference phenomenon and is further reduced by the MRR followed. Therefore, the output optical power dependent solely upon the power of the optical first-order sidebands, which depends on the interference result being determined by the phase difference and amplitude magnitude of the reference signal and unknown signal. It is crucial to note that at least two reference signals with different phases must be loaded forward and backward on one arm of the DDMZM for the measurement of a single unknown signal. By detecting the photocurrent DC when the phase of the reference signal is different, a system of equations can be solved to estimate the phase difference and amplitude ratio between the unknown signal and the reference signal, thus realizing the phase amplitude measurement of the microwave signal. The exact process of solving the equations and more detail for the operation principle of the microwave phase and can be seen in Note S2, Supporting Information.

**Design and characterization**

As shown in Figure 1(d), the microwave photonic measurement system is implemented on the silicon nitride ($Si_3N_4$)-loaded TFLN platform where the $Si_3N_4$ layer instead of the TFLN layer is etched to form a $Si_3N_4$-TFLN hybrid waveguide.[45-46] The thickness of TFLN and $Si_3N_4$ layers are both 0.3 μm, and the thickness of the buried oxide, the gold electrode and the Ti micro-heater is 4.7, 0.5, and 0.1 μm, respectively. The fabrication of coplanar waveguide electrodes is undertaken with the objective of constructing the travelling wave electrodes. The design of travelling wave electrodes comprises three key aspects as follow: (i) the index match between the RF signal and the optical signal, (ii) the RF impedance match, and (iii) the relatively low RF loss.[47] In consideration of the aforementioned trade-offs, the widths of signal and ground electrodes are chosen as 50 μm and 100 μm, respectively and the gap between signal and ground electrodes is designed to be 6 μm, which ensures a high modulation efficiency of the modulator. The radius of the MRR is designed to be 300 μm with a waveguide width of 1 μm and a gap of 0.85 μm between the bus waveguide and MRR, achieving the high-quality factor and desired FSR. Considering the trade-off between the roll-off and the frequency range of the ACF, the difference between the lengths of the two arms of the AMZI discriminator is designed as 900 μm to achieve the suitable FSR. It is worth noting that although the group refractive index of the waveguide differs in the crystallographic X and Z direction due to the anisotropic nature of the lithium niobate, the difference is too small to affect the selection of the two interfering arm length difference of the AMZI (see Note S1, Supporting Information for more detail).

The performance of the key building block (modulator, MRR, AMZI) is shown in Figure 1e-g, respectively. With an electro-optical bandwidth exceeding 80 GHz at a reference frequency of 1 GHz (Figure 1e), the modulator is capable of converting high-frequency-band microwave signals into the optical domain, providing the foundation for the ultra-high bandwidth microwave photonic measurement system. The extinction ratio, FSR and the Q factor of the MRR are 19 dB, 87GHz and $1.3 \times 10^5$, respectively (Figure 1f), ensuring that the MRR can further effectively suppress the carrier without affecting the first-order sidebands. The extinction ratios of two output ports of the AMZI with FSR of 142 GHz are 13 dB and 11 dB, respectively (Figure 1g). The disparate extinction ratios observed at the two AMZI ports can be attributed primarily to the unequal MMI splitting resulting from the discrepancy in the manufacturing process, which has a minor impact on the frequency measurement module (see Note S2, Supporting Information for more detail). Additional resonance peaks attributable to the preceding MRR are discernible in the AMZI transmission spectral curve and the low resolution (20 pm) of the optical spectrum analyzer (YOKOGAWA AQ6370D) results in low measured extinction ratios for these resonance peaks.

**Experiments and results**

**Microwave frequency measurement**

The experimental setup for microwave frequency measurement is illustrated in Figure 2a. A continuous wave optical carrier emitted from a tunable laser source (TLS, Santec TSL-570) is firstly coupled into the polarization controller (PC) and amplified by an erbium-doped fiber amplifier (EDFA, KEOPSYS CEFA-C-PB-LP-SM-23-MSA1-B201-FA-FA) before being coupled to the chip by an on-chip grating coupler. The microwave signals generated by an arbitrary waveform generator (AWG, Keysight M8199B 256 GSa/s) are combined with the bias voltage provided by a multichannel power supply (MPS, GWINSTEK GDP-3303S) with the aid of a bias tee (SHF BT65R) and fed into the MZM by a high-speed 50 Ω ground-signal-ground (GSG) probe (T-PLUS 110 GHz).The optical output signal is amplified by the EDFA in order to compensate for the on-chip insertion loss and then detected by a photodetector (PD, Finisar XPDV3120R 70 GHz). A low-speed oscilloscope (Tektronix TDS2012C) with a operational bandwidth of 100 MHz is employed to capture the DC component and low-frequency noise of the electrical signals

In advance of operation, microwave frequency measurement systems require ACF curves to be obtained for the purpose of calibration. A linear chirp frequency (LCF) signal with a frequency range of 5–65 GHz, a pulse width of 3 μs, and a repeat interval of 4 μs is input to the microwave frequency measurement module. The ratio of the two output power responses collected by the oscilloscope is subjected to a polynomial regression process in order to obtain the ACF, as shown in Figure 2b. The ACF curve obtained is in turn utilized to estimate the frequency of the unknown microwave signals. From Figure 2c-d that illustrate the estimated-input frequency relationship and the frequency measurement error, respectively, it can be seen that the microwave frequency measurement module operates over a bandwidth of up to 65 GHz, with a root-mean-square (RMS) measurement error of 450 MHz. The amplified spontaneous noise introduced by the EDFA represents the primary source of the measurement error, which can be substantially mitigated by obviating the necessity of the EDFA through the integration of additional devices, such as laser sources and PDs, on a single chip.

We experimentally demonstrate the ability of the microwave frequency measurement module to perform dynamic identification of several typical frequency-varying RF signals, including LCF, HF, and QCF signals. Figure 2e illustrates the normalized voltage as well as the frequency of the input microwave signal (part) with the repetition interval of 4 μs plotted against time. The chirping frequency ranges of the LCF signals with the pulse width of 3 μs and the hopping frequency ranges of the HF signals with the hopping time interval of 0.8 μs are selected to correspond to the X to Ku-band (8–18 GHz), K-band (18–27 GHz), Ka-band (27–40 GHz), and U-band (40–60 GHz) frequency ranges, respectively. The chirping frequency of the QCF signals ranges from 10 to 20 GHz, 20 to 30 GHz, 30 to 40 GHz, and 40 to 50 GHz (Q-band), respectively. As shown in Figure 2g-i, the instantaneous frequency profiles measured are, in general, in accordance with those of the input ones, demonstrating the ability of the frequency measurement module to achieve the real-time frequency identification with the ultra-high operating bandwidth (up to 60 GHz).

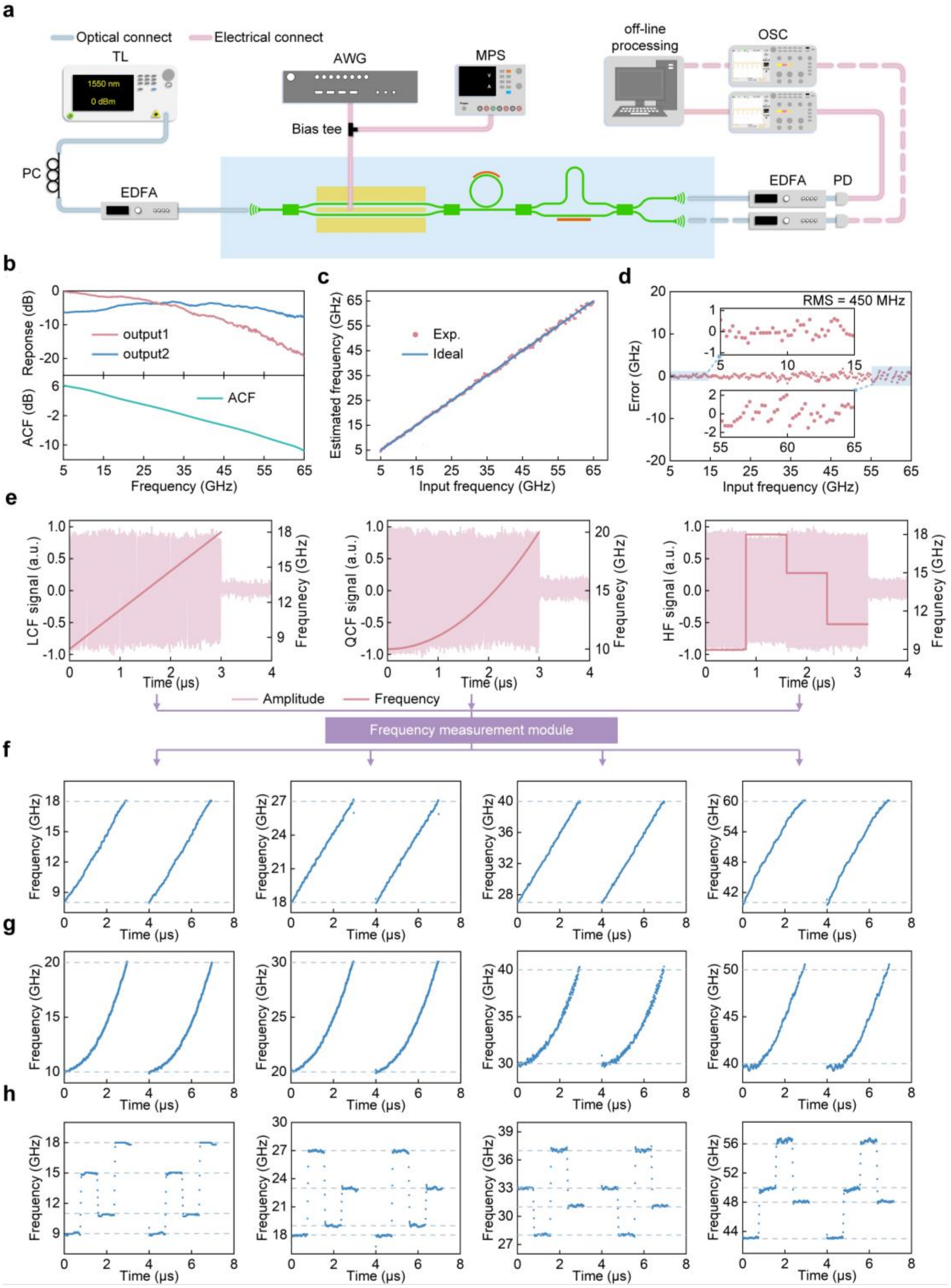

**Figure 2. Microwave frequency measurement. a.** Experimental setup for the microwave frequency measurement. **b.** The electrical responses of two output (upward) and the corresponding ACF curve (below) **c.** Frequency estimation measurement over 5–65 GHz. **d.** Frequency estimation errors over 5-65 GHz with an RMS of 450MHz. Inset shows enlarged estimation results at the range of 5–15 GHz and 55–65 GHz. **e.** Input dynamically frequency-varying microwave signals, including HF signal (left), LCF signal (middle) and QCF signal (right). **f-g.** Real-time frequency readout of HF signal (f), LCF signal (g)

and QCF signal **h.** AWG: arbitrary waveform generator, MPS: multichannel power supply, TL: tunable laser, PC: polarization controller, EDFA: erbium-doped fiber amplifier, PD: photodetector, OSC: oscilloscope, ACF: amplitude comparison function, RMS: root-mean-square, LCF: linear chirp frequency, QCF: quadratic chirp frequency, HF: hopping frequency.

## Microwave phase and amplitude measurement

The experimental setup of the amplitude-phase measurement module is essentially analogous to that of the frequency measurement module as illustrated in Figure 3a. The bias tee is not employed as the operating point of the DDMZM is controlled through the direct application of a voltage to the microheater situated on one of its arms. Furthermore, the GSGSG probe (GGB 110 GHz) is a mandatory component due to the requirement to load different high-speed signals (the unknown signal and the reference signal) on each of the two arms of the DDMZM. It is worth noting that the dashed lines in Figure 3a represents the use of multiple input reference signals of varying phases to facilitate the measurement of a single unknown signal.

Initially, phase and amplitude measurements are conducted on the signals with a peak-to-peak voltage of 2.7V over a frequency range of 10 to 60 GHz in 10 GHz increments, and over a phase range of 0 to 330° in 30° increments, whose results are presented in Figure 3b. As illustrated in Figure 3c, the distribution of measurement errors at varying frequencies reveals that, although the measurement error increases with higher frequency bands, the RMS of phase and amplitude at 60 GHz remains below 2° and 3%, respectively.

Furthermore, the phases and the amplitudes of the unknown signals at frequencies of 15GHz and 30GHz are measured and the results are illustrated in Figure 3d. The phase difference varies between 0° and 90° in increments of 22.5°, the peak-to-peak voltage of the reference signal is set at 1.8 V, and the peak-to-peak voltages of the unknown signals are selected as 0.9, 1.35, 1.8, 2.25, and 2.7 V, respectively. It is worth emphasizing that the amplitude non-linear response presented of the link and the electronics, especially the AWG, leads to a discrepancy between the amplitude ratio of the microwave signals loaded on the electrodes of the two arms of the DDMZM and the setup, which ultimately results in the amplitude measurements deviating from the preset values to a certain extent. By compensating for the nonlinear response (blue and red dashed lines in Figure 3d indicate the set value and the compensated value of amplitude, respectively), the RMS of the percentage deviation of the amplitude measurements is reduced from 3.43% to 1.09% as shown in Figure 3e.

In addition, the capacity of the phase amplitude measurement module to identify phase hopping, amplitude hopping and the phase-amplitude hopping signals is illustrated. Figure 3g illustrates the evolution of the calibrated amplitude ratio and phase difference between the input unknown signal and the reference signal over time. The repetition interval and the hopping time interval of the phase/amplitude/phase-amplitude hopping signals with the frequency of 30 GHz are chosen as 4 μs and 1 μs, respectively For the hopping phase signal, the peak-to-peak voltage is set to a value (2.7V) the same as that of the reference signals, and the phase sequences hop in the following order: 0°, 60°, 90° and 30°. For the amplitude hopping signal and phase-amplitude hopping signal, the hopping phase difference over time ranges from 0° to 90° with an interval of 22.5° and the set values of peak-to-peak voltage hop at 0.9, 1.35, 1.8, 2.25, and 2.7 V (the peak-to-peak voltage of the reference signal is set as 1.8 V). As illustrated in Figure 3h, the real-time phase and amplitude readout processed through the microwave phase and amplitude measurement module, exhibits a general correlation with the input values. It can be noted that a comparable level of noise causes a larger jump in the phase/amplitude reading for a lower power signal compared to a higher power signal. (see Note S4, Supporting Information for more detail)

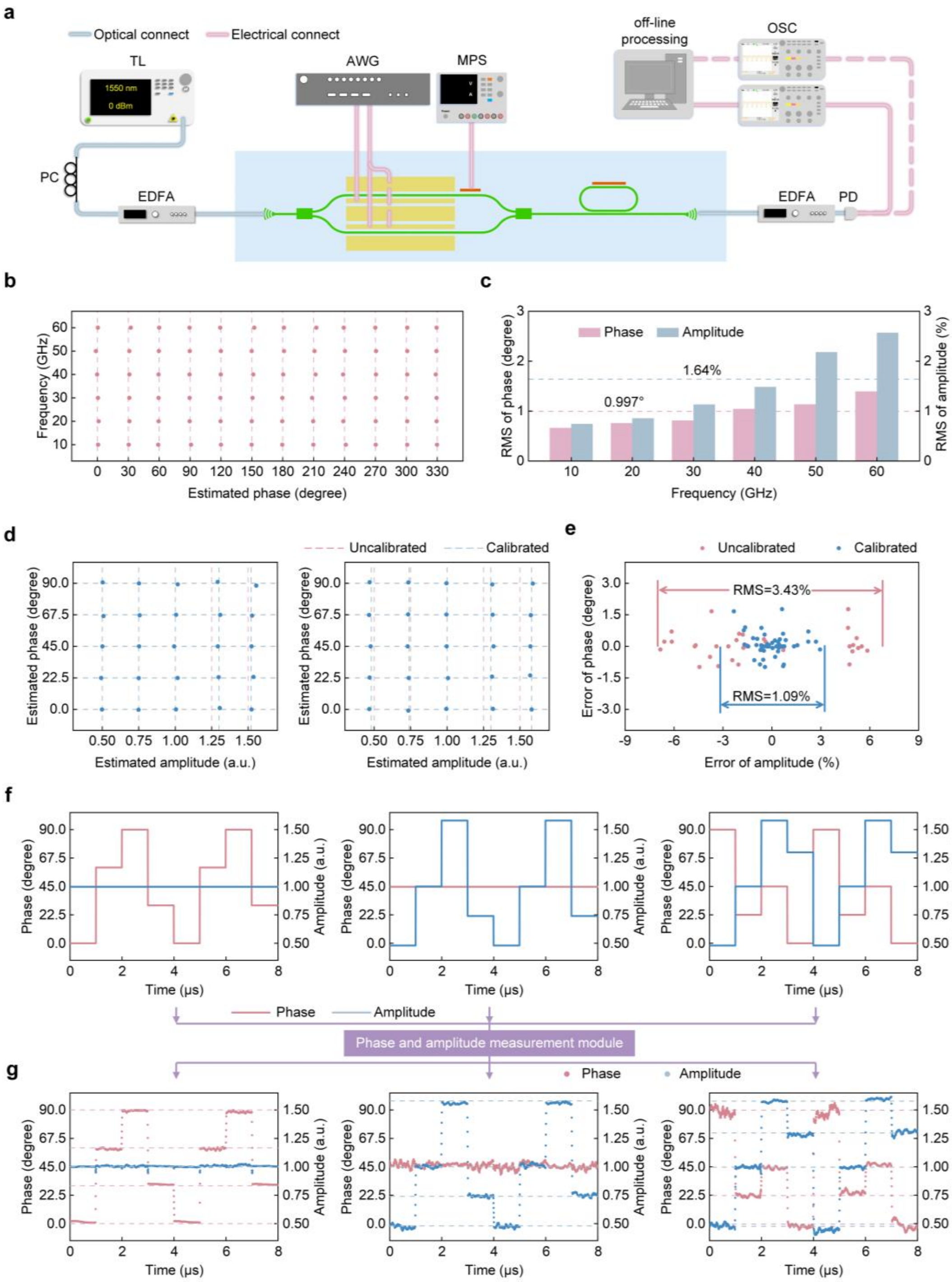

**Figure 3. Microwave phase amplitude measurement. a.** Experimental setup for the phase amplitude measurement. **b.** Phase and amplitude estimation measurement for the signals with a peak-to-peak voltage of 2.7V over a frequency range of 10 to 60 GHz in 10 GHz increments, and over a phase range of 0° to 360° in 30° increments **c.** The statistical distribution of RMS error for phase amplitude measurements at varying frequency. **d.** Phase/amplitude estimation measurement for the signals with the frequency of 15 GHz (left) and 30 GHz (right). The measurement is conducted over a phase range of 0° to 360° in 30° increments and over a peak-to-peak voltage range of 0.9, 1.35, 1.8, 2.25, and 2.7 V. **e.** The distribution of measurement errors prior to (red) and following (blue) calibration. **f.** Input dynamically phase/amplitude-varying microwave signals, including HP signal (left), HA signal (middle) and phase-amplitude hopping signal (right). **g.** Real-time phase/amplitude readout of HP signal (left), HA signal (middle) and phase-amplitude hopping signal (right). AWG: arbitrary waveform generator, MPS: multichannel power supply, TL: tunable laser, PC: polarization controller, EDFA: erbium-doped fiber amplifier, PD: photodetector, OSC: oscilloscope, RMS: root-mean-square, HP: hopping phase, HA: hopping amplitude.

## Reconstruction of sinusoidal signal

The joint utilization of the two modules enables the reconstruction of the time-domain information of an unknown sinusoidal microwave signal, as all parameters (frequency, phase and amplitude) of its time-domain feature are measured. The time-domain reconstruction of a sinusoidal signal in the ideal state can be achieved through the following steps: (1) measure the frequency of the unknown signal; (2) determine the reference signal; (3) measure the phase difference and the amplitude ratio of the unknown signal with the reference signal; and (4) reconstruct the signal. However, the aforementioned process could not be actualized for the following reasons. The FTPM based microwave frequency measurement module suffers from unavoidable errors in measuring the signal frequency, resulting in the failure to lock the frequency of the unknown signal with complete accuracy. In the event of a discrepancy between the frequency of the unknown signal and that of the reference signals, the optical first-order sidebands of the modulated signals on the two arms of the DDMZM do not interfere with each other after passing through the MMI, which means that the DC photocurrent power detected by the PD is no longer correlated with the phase difference between the unknown signal and the reference signal.

Nevertheless, the electrical signals generated by PD beating the two optical first-order sidebands modulated by the unknown signal and the reference signal can be captured by the oscilloscope and subsequently analyzed to obtain the frequency difference between the input unknown signal and the reference signal, provided that its value is less than the operational bandwidth of the oscilloscope, as illustrated in Figure 4a (see Note S4, Supporting Information for more detail).In light of the aforementioned considerations, the reconstruction of sinusoidal signals in the actual state can be achieved by the following steps, as illustrated in Figure 4b: (1) measure the frequency of the unknown signal; (2) determine the reference signal; (3) correct of the frequency of the reference signal; (4) measure the phase difference and the amplitude ratio of the unknown signal with the reference signal; and (5) reconstruct the signal. Step (3) serves to compensate for the inherent measurement errors of the microwave frequency measurement module, while simultaneously constituting a necessary condition for the normal operation of the microwave phase amplitude measurement module.

A sinusoidal signal with the frequency of 30 GHz, the set peak-to-peak voltage of 2.7 V, and the phase difference of 67.5 degrees from the reference signal (with the set peak-to-peak voltage of 1.8 V) is reconstructed through the aforementioned steps. The input signals and output beat signals for step (3) are shown in Figures 4b-c, respectively. The reconstructed signal is in close alignment with the ideal value, as illustrated in Figure 4e (the time deviation of 22 fs from the ideal value is equivalent to a phase deviation from the ideal value of 0.24°). By reducing the maximum measurement error of the microwave frequency measurement module, or by utilizing oscilloscopes with a higher operating bandwidth, it is possible to reconstruct the time-domain information for sinusoidal microwave signals in a wider frequency band, or even for more complex signals such as chirp frequency signals and hopping frequency signals.

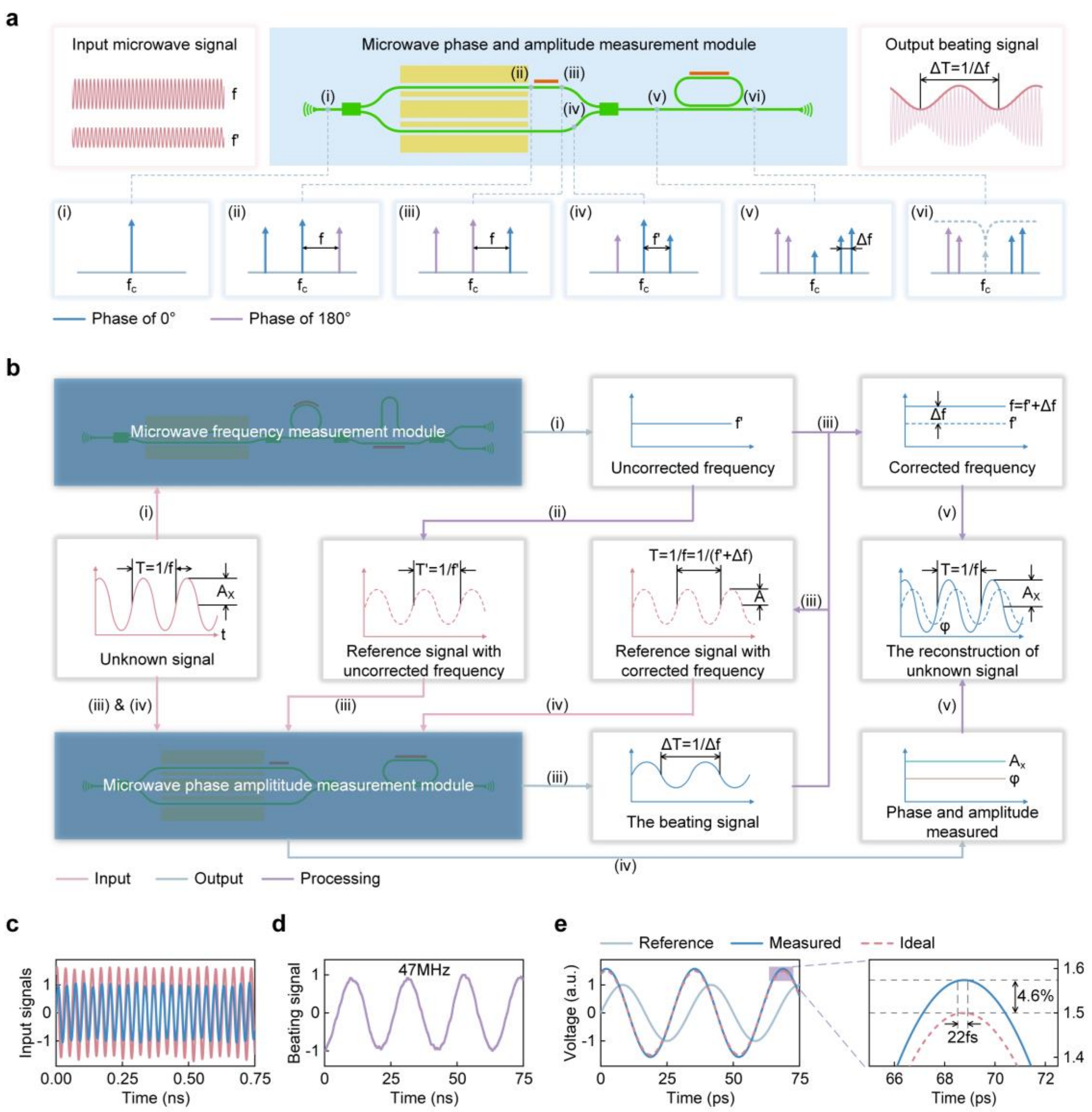

**Figure 4. Time-domain reconstruction of sinusoidal microwave signal a.** The optical frequency-domain signal schematic spectra (blue) at different locations of the chip and the electrical voltage-time schematic diagrams (pink), when the frequencies of the two input signals to the microwave phase amplitude measurement module are disparate. **b.** Flow diagram of time-domain reconstruction of sinusoidal microwave signal. Steps are as followed: (i) measure the frequency of the unknown signal; (ii) determine the reference signal; (iii) correct of the frequency of the reference signal; (iv) measure the phase difference and the amplitude ratio of the unknown signal with the reference signal; and (x) reconstruct the signal. **c-d.** The input signals (c) and output beating signal (d) for correction of the reference signal. The two signals, with frequencies of 30 GHz (red) and 30.047 GHz (blue), and peak-to-peak voltages of 2.7 V (red) and 1.8 V (blue), are input to the microwave measurement module, respectively. **e.** Time-domain reconstruction result for a signal with the frequency of 30 GHz with the frequency of 30 GHz, the set peak-to-peak voltage of 2.7 V, and the phase difference of 67.5 degrees

from the reference signal (with the set peak-to-peak voltage of 1.8 V).

## DISSCUSSION

In summary, we have designed, fabricated, and demonstrated an ultra-bandwidth TFLN-based integrated microwave photonic multi-parameter measurement system that is capable of measuring the frequency, phase and amplitude of the RF signals. By employing the TFLN platform that exhibits superior modulation performance, we achieve the measurements of the frequency, amplitude and phase of the microwave signals with an ultra-high (up to 60 GHz) bandwidth and with a low RMS error (450MHz, 3.43° and 1.64% of the measurement for frequency, phase and amplitude, respectively). Furthermore, we illustrate the ability of system to discern the varieties of microwave signals, encompassing CF, HF, HP, HA and hopping phase-amplitude signals. Finally, in order to present the versatility of the system, we demonstrate the time-domain reconstruction of a sinusoidal microwave signal experimentally, as all parameters (frequency, phase and amplitude) of its time-domain feature are measured.

Predictably, integrated microwave photonics frequency measurement systems with enhanced operational bandwidths, reduced measurement errors and expanded functionality could be achieved through the optimization of the electro-optical modulator structure, the improvement of the manufacturing process and the refinement of the measurement implementation scheme. Moreover, other components of the microwave photonics measurement system, including low-noise laser sources[48-50], high-response photodetectors, and highly flexible electronic integrated circuits, are anticipated to be assembled on the TFLN platform through heterogeneous integration schemes, providing a highly compact, cost-effective, and high-performance integrated microwave signal measurement system for next generation information and communication fields.[51-52]

## References

1. Ostro, Steven J. "Planetary radar astronomy." Reviews of Modern Physics 65.4 (1993): 1235.

2. Ilderem, Vida. "The technology underpinning 5G." Nature Electronics 3.1

(2020): 5-6.
3. Elbert, Bruce R. Introduction to satellite communication. Artech house, 2008.
4. Hecht, Jeff. "The bandwidth bottleneck." Nature 536.7615 (2016): 139-142.
5. Boes A, Chang L, Langrock C, et al. Lithium niobate photonics: Unlocking the electromagnetic spectrum[J]. Science, 2023, 379(6627): eabj4396.
6. Zou, Jianping Yao. "Photonics for microwave measurements." Laser & Photonics Reviews 10, no. 5 (2016): 711-734.
7. Capmany, José, and Dalma Novak. "Microwave photonics combines two worlds." Nature photonics 1.6 (2007): 319.
8. Pan, Shilong, and Jianping Yao. "Photonics-based broadband microwave measurement." Journal of Lightwave Technology 35.16 (2016): 3498-3513.
9. Fandiño, Javier S., and Pascual Muñoz. "Photonics-based microwave frequency measurement using a double-sideband suppressed-carrier modulation and an InP integrated ring-assisted Mach–Zehnder interferometer filter." Optics letters 38.21 (2013): 4316-4319.
10. Pagani, Mattia, et al. "Low-error and broadband microwave frequency measurement in a silicon chip." Optica 2.8 (2015): 751-756.
11. Burla, Maurizio, et al. "Wideband dynamic microwave frequency identification system using a low-power ultracompact silicon photonic chip." Nature communications 7.1 (2016): 13004.
12. Jiang, Hengyun, et al. "Wide-range, high-precision multiple microwave frequency measurement using a chip-based photonic Brillouin filter." Optica 3.1 (2016): 30-34.
13. Zhao, Chang-Xin, et al. "Multiscale construction of bifunctional electrocatalysts for long-lifespan rechargeable zinc–air batteries." Advanced Functional Materials 30.36 (2020): 2003619.
14. Zhu, Beibei, et al. "High-sensitivity instantaneous microwave frequency measurement based on a silicon photonic integrated Fano resonator." Journal of Lightwave Technology 37.11 (2019): 2527-2533.
15. Tao, Yuansheng, et al. "Fully on-chip microwave photonic instantaneous

frequency measurement system." Laser & Photonics Reviews 16.11 (2022): 2200158.

16. Zhao, Mengyao, et al. "Photonic-Assisted Microwave Frequency Measurement Using High Q-Factor Microdisk with High Accuracy." Photonics. Vol. 10. No. 7. MDPI, 2023.

17. Wang, LiHeng, et al. "Integrated Ultra-Wideband Dynamic Microwave Frequency Identification System in Lithium Niobate on Insulator." Laser & Photonics Reviews: 2400332.

18. Volyanskiy, Kirill, et al. "Applications of the optical fiber to the generation and measurement of low-phase-noise microwave signals." JOSA B 25.12 (2008): 2140-2150.

19. Zhu, Dengjian, et al. "Wideband phase noise measurement using a multifunctional microwave photonic processor." IEEE Photonics Technology Letters 26.24 (2014): 2434-2437.

20. Chen, Hao, and Erwin HW Chan. "Photonics-based CW/pulsed microwave signal AOA measurement system." Journal of Lightwave Technology 38.8 (2020): 2292-2298.

21. Zhao, Jianing, Zhenzhou Tang, and Shilong Pan. "Photonic approach for simultaneous measurement of microwave DFS and AOA." Applied Optics 60.16 (2021): 4622-4626.

22. Ding, Jiewen, et al. "Simultaneous angle-of-arrival and frequency measurement system based on microwave photonics." Journal of Lightwave Technology 41.9 (2023): 2613-2622.

23. Zhang, Bowen, Xiangchuan Wang, and Shilong Pan. "Photonics-based instantaneous multi-parameter measurement of a linear frequency modulation microwave signal." Journal of Lightwave Technology 36.13 (2018): 2589-2596.

24. Wang, Yi, et al. "Photonic architecture for remote multi-parameter measurement and transmission of microwave signals." Optics Express 32.10 (2024): 18033-18043.

25. Boes, Andreas, et al. "Status and potential of lithium niobate on insulator (LNOI)

for photonic integrated circuits." Laser & Photonics Reviews 12.4 (2018): 1700256.

26. Zhang, Mian, et al. "Integrated lithium niobate electro-optic modulators: when performance meets scalability." Optica 8.5 (2021): 652-667.
27. Zhu, Di, et al. "Integrated photonics on thin-film lithium niobate." Advances in Optics and Photonics 13.2 (2021): 242-352.
28. Tian, Yonghui, et al. "Integrated ultra-wideband tunable Fourier domain mode-locked optoelectronic oscillator." (2024).
29. Sun, Wenzhao, et al. "Wafer-scale thin-film lithium niobate device fabrication and characterization." TENCON 2022-2022 IEEE Region 10 Conference (TENCON). IEEE, 2022.
30. Luke, Kevin, et al. "Wafer-scale low-loss lithium niobate photonic integrated circuits." Optics Express 28.17 (2020): 24452-24458.
31. Han, Xu, et al. "Single-step etched grating couplers for silicon nitride loaded lithium niobate on insulator platform." Apl Photonics 6.8 (2021).
32. Liao H, Chen L, Zhou X, et al. Photonic Metamaterial-Inspired Polarization Manipulating Devices on Etchless Thin Film Lithium Niobate Platform[J]. Laser & Photonics Reviews, 2400381.
33. Chen L, Han X, Zhou X, et al. Demonstration of a High-Performance 3 dB Power Splitter in Silicon Nitride Loaded Lithium Niobate on Insulator[J]. Laser & Photonics Reviews, 2023, 17(11): 2300377.
34. Zhou X, Chen L, Liao H, et al. Integrated Arbitrary Multimode TM-Pass Polarizer Based on Anisotropic Optical Manipulation[J]. Laser & Photonics Reviews, 2400932.
35. Zhang, Mian, et al. "Monolithic ultra-high-Q lithium niobate microring resonator." Optica 4.12 (2017): 1536-1537.
36. Zhuang, Rong, et al. "High-Q Thin-Film Lithium Niobate Microrings Fabricated with Wet Etching." Advanced Materials 35.3 (2023): 2208113.
37. Ma, Mingyang, et al. "Multimode waveguide bends in lithium niobate on insulator." Laser & Photonics Reviews 17.5 (2023): 2200862.

38. Zhao, Weike, et al. "High-Performance Mode-Multiplexing Device with Anisotropic Lithium-Niobate-on-Insulator Waveguides." Laser & Photonics Reviews 17.5 (2023): 2200774.
39. Han, Xu, et al. "Subwavelength Grating-Assisted Contra-Directional Couplers in Lithium Niobate on Insulator." Laser & Photonics Reviews 17.10 (2023): 2300203.
40. Han, Xu, et al. "Mode and polarization-division multiplexing based on silicon nitride loaded lithium niobate on insulator platform." Laser & Photonics Reviews 16.1 (2022): 2100529.
41. Yuan, Mingrui, et al. "Integrated lithium niobate polarization beam splitter based on a photonic-crystal-assisted multimode interference coupler." Optics Letters 48.1 (2022): 171-174.
42. Feng, Hanke, et al. "Integrated lithium niobate microwave photonic processing engine." Nature 627.8002 (2024): 80-87.
43. Li Z, Wang R N, Lihachev G, et al. High density lithium niobate photonic integrated circuits[J]. Nature Communications, 2023, 14(1): 4856.
44. Luke K, Kharel P, Reimer C, et al. Wafer-scale low-loss lithium niobate photonic integrated circuits[J]. Optics Express, 2020, 28(17): 24452-24458.
45. Zhang, Pu, et al. "High-speed electro-optic modulator based on silicon nitride loaded lithium niobate on an insulator platform." Optics letters 46.23 (2021): 5986-5989.
46. Han, Xu, et al. "Integrated photonics on the dielectrically loaded lithium niobate on insulator platform." JOSA B 40.5 (2023): D26-D37.
47. Renaud, Dylan, et al. "Sub-1 Volt and high-bandwidth visible to near-infrared electro-optic modulators." Nature Communications 14.1 (2023): 1496.
48. Luo, Qiang, et al. "Advances in lithium niobate thin-film lasers and amplifiers: a review." Advanced Photonics 5.3 (2023): 034002-034002.
49. Ye, Kaixuan, et al. "Surface acoustic wave stimulated Brillouin scattering in thin-film lithium niobate waveguides." arxiv preprint arxiv:2311.14697 (2023).
50. Rodrigues, Caique C., et al. "On-chip backward stimulated Brillouin scattering

in lithium niobate waveguides." CLEO: Science and Innovations. Optica Publishing Group, 2024.

51. Perez, Daniel, et al. "Silicon photonics rectangular universal interferometer." Laser & Photonics Reviews 11.6 (2017): 1700219.

52. Wen, He, et al. "Few-mode fibre-optic microwave photonic links." Light: Science & Applications 6.8 (2017): e17021-e17021.

**Acknowledgments**

The authors acknowledge the facilities, and the scientific and technical assistance, of the Micro Nano Research Facility (MNRF) and the Australian Microscopy & Microanalysis Research Facility at RMIT University. This work was performed in part at the Melbourne Centre for Nanofabrication (MCN) in the Victorian Node of the Australian National Fabrication Facility (ANFF).

**Funding**

This work was supported by National Natural Science Foundation of China (NSFC) (62075091, 62205135), Key Research and Development of Gansu Province (22YF7GA008, 23YFGA0007), Natural Science Foundation of Gansu Province (23JRRA1026, 22JR5RA493), and Australian Research Council (ARC) grant (DP190102773).

**Conflict of Interest**

The authors declare no conflict of interest.

**Data Availability Statement**

All the data supporting the findings in this study are available in the paper and Supplementary Information. Additional data related to this paper are available from the corresponding authors upon request.